\documentclass[twocolumn,floatfix,pra,aps,showpacs]{revtex4}
\usepackage{epsfig,graphicx,tabularx,amsmath}
\usepackage{color}

\begin{document}
\title{Dipolar condensates confined in a toroidal trap: ground state and vortices}

\author{M. Abad}
\affiliation{Departament d'Estructura i Constituents de la Mat\`{e}ria,\\
Facultat de F\'{\i}sica, Universitat de Barcelona, E--08028 Barcelona, Spain}
\author{M. Guilleumas}
\affiliation{Departament d'Estructura i Constituents de la Mat\`{e}ria,\\
Facultat de F\'{\i}sica, Universitat de Barcelona, E--08028 Barcelona, Spain}
\author{R. Mayol}
\affiliation{Departament d'Estructura i Constituents de la Mat\`{e}ria,\\
Facultat de F\'{\i}sica, Universitat de Barcelona, E--08028 Barcelona, Spain}
\author{M. Pi}
\affiliation{Departament d'Estructura i Constituents de la Mat\`{e}ria,\\
Facultat de F\'{\i}sica, Universitat de Barcelona, E--08028 Barcelona, Spain}
\author{ D. M. Jezek   }
\affiliation{Departamento de F\'{\i}sica, Facultad de Ciencias Exactas
y Naturales,
Universidad de Buenos Aires, RA-1428 Buenos Aires, Argentina}
\affiliation{Consejo Nacional de Investigaciones Cient\'{\i}ficas y
T\'ecnicas, Argentina}

\date{\today}
\begin{abstract}
We study a  Bose-Einstein condensate of
$^{52}$Cr atoms confined in a toroidal trap with a variable strength of
$s$-wave contact interactions. 
We analyze the effects of the anisotropic nature of the dipolar interaction by considering the magnetization axis to be perpendicular to the trap symmetry axis. 
In the absence of a central repulsive barrier, when the trap is purely harmonic, the effect of reducing the scattering length is a tuning of the geometry of the system: from a pancake-shaped condensate when it is large, to a cigar-shaped condensate for small scattering lengths. 
For a condensate in a toroidal trap, the interaction in combination with the central repulsive Gaussian barrier
produces an azimuthal dependence of the particle density for a fixed radial distance.
We find that along the magnetization direction
the density decreases as the 
scattering length is reduced but presents two symmetric density peaks in the perpendicular axis.
For even lower values of the scattering length we observe that the system 
 undergoes a dipolar-induced symmetry breaking phenomenon.
The whole density becomes concentrated in one of the peaks,
resembling an origin-displaced cigar-shaped condensate.
In this context we also analyze stationary vortex states and their associated velocity field, finding that this latter also shows a strong azimuthal dependence for small scattering lengths. 
The expectation value of the angular momentum along the $z$ direction provides a qualitative measure of the difference between the velocity in the different density peaks.

\end{abstract}
\maketitle
\section{Introduction}\label{Intro}
Due to the large magnetic dipole moment that chromium atoms  possess, 
a gas of ultracold $^{52}$Cr is a very suitable system for
studying the effects of the dipolar interaction in different types of condensates.
Although chromium condensates show both dipolar as well as $s$-wave contact interactions, a possible way to enhance the effects of the former is to reduce the value of the scattering length, which can be achieved with Feshbach resonances \cite{Werner2005}. The wide range of configurations attainable in dipolar condensates have fostered an intense theoretical and experimental activity in the field (see Ref.~\cite{Lahaye2009} for a recent review). There have been studies delving into the anisotropic character of the interactions \cite{Stuhler2005, Giovanazzi2006}, the sensitivity of the system to the trap aspect ratio \cite{Santos2000, Koch2008}, the Thomas-Fermi limit in dipolar condensates \cite{Eberlein2005}, the effects of long-range in the collective excitation spectrum \cite{Goral2002, Yi2002, Dell2004, Ronen2006}, the appearance of a roton minimum \cite{Santos2003, Ronen2007}, and the instability and collapse of the system \cite{Wilson2009a, Bijnen2007, Lahaye2008}, among others.

The experimental realization of  dipolar Bose-Einstein 
condensates (BECs) in harmonic traps \cite{gri05, Beaufils2008} 
opens the possibility of studying their properties in more complex confining potentials. Optical lattices have been already addressed in the literature (see the review \cite{Lahaye2009} and references therein), as well as double-well potentials \cite{xi09,Asad2009}, but the present work is to our knowledge the first where dipolar condensates are studied in toroidal traps. Experimentally, toroidal traps can be realized by shining the condensate with a blue-detuned laser beam \cite{ryu07}.

Spontaneous symmetry breaking in quantum gases has recently attracted a lot of interest, especially in double-well confining potentials. 
For contact interaction condensates, the authors in Ref.~\cite{may08} have predicted
that this phenomenon can be produced
by an external inhomogeneous modulation of the scattering length.
Within the framework of dipolar condensates, symmetry breaking phenomena have also been predicted to appear \cite{xi09,Asad2009}. In Ref.~\cite{xi09}, the authors have
observed that 
depending on the magnetization direction the system undergoes a symmetry
breaking, while in Ref.~\cite{Asad2009} it is the strength of the dipolar interaction that drives the same kind of phenomenon. In all these cases the symmetry broken configuration is related to the self-trapped state of a Josephson junction.

In this work we are interested  in 
determining the physical
consequences of the combination of long range anisotropic interactions with a ring-shaped condensate.
We concentrate in the particular  situation in which
the magnetization direction is perpendicular to the trap symmetry axis.
Due to the anisotropic character of the dipolar interaction, the particle density does not 
conserve the  azimuthal symmetry of
the confining potential, but presents two well defined peaks in the direction perpendicular to the magnetization. 
We find that when the scattering length is reduced, starting from  the  chromium natural one, 
the height of these peaks increases.
Moreover, for a small enough value one of them disappears, 
giving rise to a symmetry breaking phenomenon.
This behavior could be experimentally confirmed
since either trapping atoms in toroidal traps \cite{ol07},
controlling the $s$-wave scattering
length \cite{Koch2008} and the condensation of dipolar gases \cite{gri05, Beaufils2008} are nowadays successfully achieved.

Another important issue that has always attracted much interest in the field is the superfluid character of BECs. The presence of quantized vortices is a clear signature of superfluidity in quantum systems \cite{don91} and this subject has received intensive research in the last decades (see \cite{Fetter2009} for a recent review). It has been recently shown that purely $s$-wave condensates in toroidal traps are also capable of sustaining (metastable) vortex states, which produce observable persistent flows \cite{ryu07}. This fact has attracted renewed interest in the physics arising in multiply connected geometries. In particular, in the last years, the formation  \cite{wei08}, stability \cite{cap09},
 dynamics \cite{mas09,cat09},
and dissipation \cite{pia09} of vortices in toroidal traps
have been investigated.

The particular characteristics of dipolar interactions in BECs have raised numerous questions regarding the physics of vortices in such systems. Strong theoretical effort has been invested into the characterization of vortices in dipolar condensates \cite{Yi2006, dell07, Klawunn, Wilson2009, abad09}. 
In this work we study stationary vortex states whose vorticity is along the
symmetry axis of the toroidal trap.
Since the dipolar interaction introduces a density angular dependence,
the velocity field of
the vortices is accordingly modified.
To gain insight into this effect, we study the velocity fields created by phase-imprinted vortex states  as a function of the scattering length, i.e. of the angular anisotropy of the condensate density.

This work is organized as follows. In Sec.~\ref{Theory} we describe the
theoretical framework and the system under study.
In Sec.~\ref{GS} we investigate the
effects on the ground state that appear from the combination of the dipolar interaction
and the contact interaction when the trap geometry is changed. 
We work  with either a harmonic  or a toroidal trap and discuss the differences between
both cases.
We tune the scattering length 
 to analyze  the case in which it is 
small enough to be comparable to the dipole-dipole interaction, and show that in the 
toroidal trap a
symmetry breaking phenomenon arises.
In Sec.~\ref{Vortex} we analyze vortex states and their associated velocity field.
Finally, a summary and concluding remarks are
offered in Sec.~\ref{Conclusions}.

\section{Theoretical framework}\label{Theory}

We consider a Bose-Einstein condensate with $N = 5 \times 10^4 $ atoms of $^{52}$Cr,
at zero temperature,
confined in a toroidal trap.
The trapping potential  $ V_{\text{trap}}(\mathbf{r}) $
is the sum of  an axially symmetric harmonic potential 
and the potential provided by a Gaussian laser beam \cite{ryu07}. The laser intensity
is proportional to $ V_0 $ and it  
propagates along the z direction with waist $  w_0 $.
Thus the trapping potential in cylindrical coordinates reads,
\begin{equation}
V_{\text{trap}}(\mathbf{r})=
  \frac{m}{2}\, (\omega_{\perp}^2 r^2 +\omega_{z}^2 z^2) \, + V_0 \, \exp ( -2 \, r^2/\; w_0^2)\, ,
\label{beam}
\end{equation}
where $r^2=x^2+y^2$, $m$ is the atomic mass, and $\omega_{\perp}$
and $\omega_{z}$ are the radial and axial angular trap frequencies,
respectively.
These  harmonic trap 
frequencies are fixed at $\omega_\perp=8.4\times2\pi$ s$^{-1}$ and
$\omega_z=92.5\times2\pi$ s$^{-1}$, giving an aspect ratio of 
$\lambda=\omega_z/\omega_\perp  = 11 $ and thus describing a pancake-shaped configuration, being $z$ the trap symmetry axis. 
The laser parameters we have used are $ V_0 =15\hbar \omega_{\perp} $ 
and  $ w_0 = 10 \mu$m.

The magnetic dipole moment of $^{52}$Cr is $\mu
= 6 \mu_B$, where $\mu_B$ is the Bohr magneton. 
 We consider that chromium atoms interact
not only via $s$-wave contact interactions but also via the
dipole-dipole interaction
\begin{equation}
v_{\text{dip}} (\mathbf{r}-\mathbf{r'})= \frac{\mu_0 \mu^2}{4 \pi}
\frac{1 - 3 \cos^2 \theta}{|\mathbf{r}-\mathbf{r'}|^3} \,,
 \label{dip-pot}
\end{equation}
where $\mu_0$ is the vacuum permeability,
$\mathbf{r}-\mathbf{r'}$ is the distance between the dipoles, and
$\theta$ is the angle between
the vector $\mathbf{r}-\mathbf{r'}$ and the dipole axis.
In this work we choose the magnetization axis to be the $y$-axis, which is perpendicular to the symmetry axis of the trap . 
 Since the dipoles are situated head to tail along $y$ direction,
the dipole-dipole interaction is  attractive along this direction and repulsive 
in the perpendicular ones.

For  weakly interacting dipolar BECs the Gross-Pitaevskii (GP) equation
\begin{eqnarray}
 & &
 \left[ -\frac{ \hbar^2}{2m} \nabla^2 + V_{\text{trap}} +
g \, |\psi(\mathbf{r})|^2 +  V_{\text{dip}}(\mathbf{r}) \right] \psi(\mathbf{r})=
 \nonumber \\
&& 
 \qquad\qquad= \tilde{\mu} \, \psi(\mathbf{r}) \,
\label{gp}
\end{eqnarray}
has proven to provide a good description of stationary states.
In this equation  $\psi(\mathbf{r})$ is the
condensate order parameter normalized to the total number of
particles  and $\tilde{\mu}$ is the chemical potential. 
The contact interaction potential is characterized by the
coupling constant $g=4\pi\hbar^2 a /m$, where $a$ is the $s$-wave scattering length. The mean-field dipolar interaction
 $V_{\text{dip}} (\mathbf{r})$ is given by
\begin{equation}
 V_{\text{dip}} (\mathbf{r})= 
\int d\mathbf{r'} v_{\text{dip}} (\mathbf{r}-\mathbf{r'}) |\psi(\mathbf{r'})|^2\ ,
\label{Vdip}
\end{equation}
which can be more easily computed by means of Fourier transform techniques (see \cite{abad09} and references therein). Inserting expression (\ref{Vdip}) into Eq.~(\ref{gp}) an integro-differential GP equation is obtained, which we solve using the imaginary-time method.

The energy density functional has the standard form but
with the additional term $E_{\text{dip}}$:
\begin{eqnarray}
 E [\psi]  && = E_{\text{kin}} + E_{\text{trap}} + E_{\text{int}} + E_{\text{dip}} 
= \nonumber\\
 && = \int\frac{ \hbar^2 }{2 m}  |\nabla \psi |^2 d {\bf r} + \int V_{\text{trap}}
 \,|\psi|^2 d {\bf r} + 
  \nonumber\\
&&+ \int\frac{g}{2} \, |\psi|^4 d {\bf r} + \frac{1}{2} \int V_{\text{dip}} \,
|\psi|^2 d\,\mathbf{r}\,.
\label{ed}
\end{eqnarray}
The last term $E_{\text{dip}}$ represents the energy
due to the dipolar interaction.

\section{Ground state}\label{GS}

In this section we analyze 
the ground state of the dipolar system,
$\psi_0(\mathbf{r})$, which is found by minimizing the total energy
(\ref{ed}), that is, by solving the GP equation (\ref{gp}).
We separately address  the cases of harmonic and harmonic plus Gaussian potential,
to better  analyze the modifications
the latter introduces.

 \subsection{Harmonic trap}

First, we analyze the effect of the dipolar interaction in a BEC confined in a harmonic trap,
 i.e. without any Gaussian barrier ($V_0=0$ in Eq.(\ref{beam})), 
and with the dipoles aligned along the $y$-axis. 
Since in this direction the 
dipolar potential is attractive, it is energetically favorable for the system to locate more atoms along the $y$-axis rather than in the $x$ one. Therefore, the net size of the condensate along the magnetization direction increases with 
respect to the purely $s$-wave case (for a more detailed discussion 
see Ref.~\cite{abad09} and references therein), whereas along $x$ and $z$ directions 
it decreases. When the dipolar effects are magnified, that is the scattering length 
is reduced, the anisotropic nature of the interaction becomes even clearer.

\begin{figure}
\epsfig{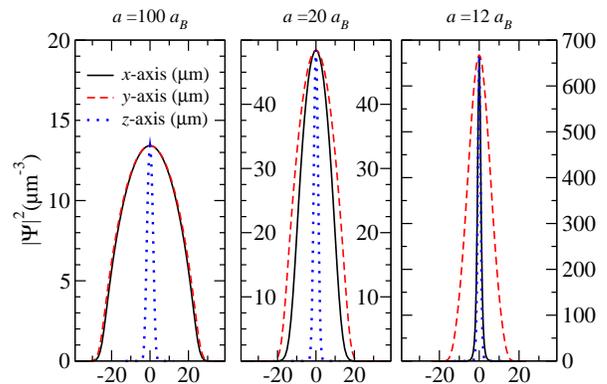} 
\caption{(Color online) Ground
state densities along $x$-axis (solid line), $y$-axis (dashed line) 
and $z$-axis (dotted line) for a harmonically trapped condensate for three different scattering lengths: $a=100, 20, 12\,a_B$ (from left to right).} \label{pancigar}
\end{figure}

From Fig.~\ref{pancigar} it can be seen that with a scattering
 length $a= 100\,a_B$ the 
 condensate is still pancake shaped, with the $x$ and $y$ sizes  
almost equal and the $z$ one
much smaller. 
This shows that for large scattering lengths the dipolar effects are almost negligible, since the dipole-dipole interaction introduces only a slight perturbation of the density and therefore the sizes in $x$ and $y$ directions are similar.
When the scattering length is reduced the dipolar effects become enhanced and the magnetization direction is privileged, yielding therefore a larger size than in the perpendicular direction. For a further reduction of the scattering length down to $ a= 12\,a_B $ the condensate
resembles a cigar with a similar size in the $x$ and $z$ directions but a
 much larger one along $y$.
For $a<12\,a_B$ the system becomes unstable. 

In order to quantitatively analyze the deformation of the condensate in the pancake trap,
 we compute the root-mean-square radius in each direction,
 namely $R_i$  where $i=x,y,z$. 
Thus for $ a=100\,a_B$ we find  $ R_x= 10.41\,\mu$m, $R_y=10.75\,\mu$m and
 $R_z=1.26\,\mu$m and thus $ R_z < R_x \simeq R_y $ 
while for $a = 12\,a_B$ we obtain $ R_x=1.36\,\mu$m, $R_y=6.13\,\mu$m and 
$R_z=0.87\,\mu$m  which verifies $R_x \simeq R_z < R_y $.
Therefore, in a BEC confined in a pancake trap potential with the dipoles
aligned perpendicularly to the trap axis, a change in the scattering length 
may be translated to a significant change in the condensate geometry.
In this particular case, just by reducing the scattering length from $100\,a_B$ to $12\,a_B$, it is possible to tune the geometry
from a nearly pancake-shaped
to a nearly cigar-shaped
 condensate.

\begin{figure}\centering
\epsfig{file=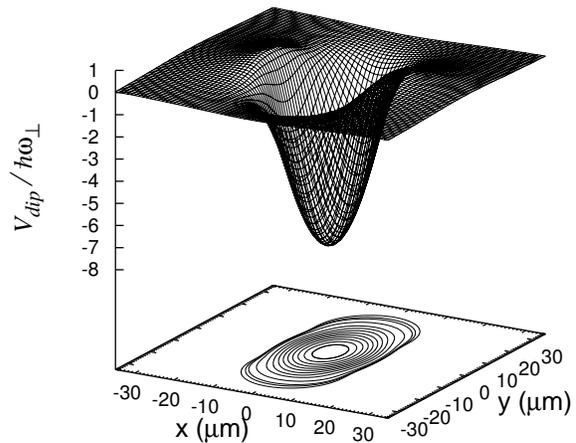,width=1.2\linewidth,clip=true}
\caption{Dipolar potential, Eq.~(\ref{Vdip}), as a function of $x$ and $y$ at $z=0$ for a harmonically trapped condensate with $ a =  20\,a_B $.}
\label{pan20n}
\end{figure}

To understand this behavior, we display
in  Fig.~\ref{pan20n}   the dipolar potential given by Eq.~(\ref{Vdip}) for $a=20\,a_B$. One can see that there exists a negative minimum at the center of the trap and that the potential
goes to zero more rapidly in the $x$ direction than in the $y$ direction. Moreover there
are two maxima around $ |x | = 16.5\, \mu$m with $ V_{max}= 1 \hbar \omega_{\perp}$.
Thus the 
energy cost to accommodate atoms along the $y$ axis is lower, 
which produces the condensate elongation in this
direction.

\subsection{ Toroidal trap }

Now we turn to the study of dipolar condensates in toroidal traps. 
The Gaussian potential in Eq.~(\ref{beam}) introduces a repulsive barrier
 along the $z$ axis. For large scattering lengths this barrier is not enough to 
burn a hole at the center of the condensate. 
However, when the scattering length is reduced the
 chemical potential of the condensate becomes smaller and eventually equals $V_0$, under which condition a hole appears at the center.
By changing $V_0$ one can choose the scattering length for the onset of toroidal geometry. 
 In what follows we will work with $V_0=15\,\hbar\omega_\perp$, 
for which a hole appears slightly above $a=30\,a_B$ (see Table \ref{tab:1}). 
We have checked that no qualitatively different behavior appears when other intensities of the Gaussian beam are used.

\begin{table}[h]
\caption{\label{tab:1} Values of the root-mean-square values and chemical
 potential for different 
scattering lengths for $ V_0 =15\, \hbar \omega_{\perp} $. 
}
\begin{ruledtabular}
\begin{tabular}{l|ccccccccc}
$a/a_B$ & $R_x (\mu m)$ & $R_y (\mu m)$ & $R_z (\mu m)$   & $\mu/\hbar\omega_\perp$ \\
\hline
$100 $ & $11.54$ & $11.78$ & $ 1.23 $  & $21.587$ \\
\hline
$30 $ & $ 9.65 $ & $ 9.94 $ & $ 1.09  $  & $ 14.903$ \\
\hline
$20  $ & $9.37$ & $9.32$ & $ 1.06 $  & $13.212$ \\
\hline
$14$  & $9.48$ & $8.45$ & $ 1.02 $  & $11.749$ \\
\hline
$12 $ (SB) & $ 10.24$ & $ 6.78$ & $ 0.89  $ &  $8.693$ \\
\end{tabular}
\end{ruledtabular}
\end{table}

\begin{figure}
\epsfig{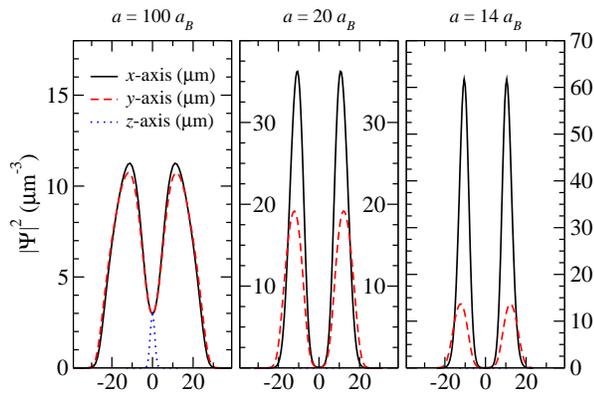}
\caption{(Color online) Density profiles of the
 ground state configuration along  $x$-axis (solid line), $y$-axis (dashed line) 
and $z$-axis (dotted line), 
for $a=100,20,14\,a_B$ (from left to right).}
\label{torus14}
\end{figure}

When the scattering length is large,
the main effect of the Gaussian barrier 
is to decrease the central density
and expand the cloud, as shown in the left panel of Fig.~\ref{torus14} for $a=100\, a_B$. In this case, the radii 
fulfill $ R_x \simeq R_y$,  with $ R_x $ slightly smaller than  $ R_y$ (see Table \ref{tab:1}). 
This behavior is similar to the previous harmonically trapped system because the dipole-dipole interaction 
introduces only a slight perturbation of the density.

When the scattering length is small, however, the situation is completely different from the purely harmonic trap.  
In this latter, the mean-field dipolar interaction was attractive along the $y$ direction and repulsive along the $x$ direction, resulting in a broadening of the condensate in the $y$ axis, namely $R_x < R_y$. In contrast, in a toroidal trap and in particular below $a=20\,a_B$, the rms-radii fulfill
$ R_x > R_y $ (see Table \ref{tab:1}), 
which certainly presents the opposite behavior 
to the purely harmonic trap system.

\begin{figure}
\epsfig{file=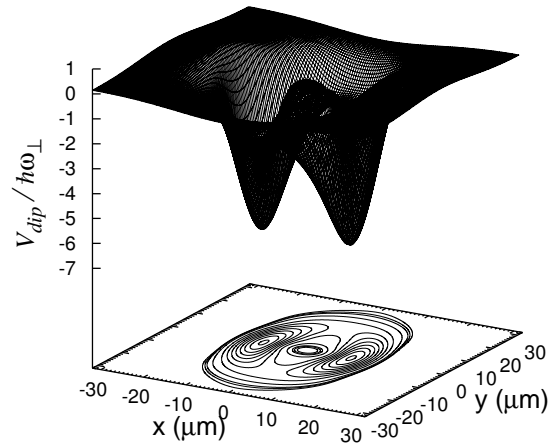,width=1.2\linewidth,angle=0,clip=true}
\caption{Dipolar potential, Eq.~(\ref{Vdip}), as a function of $x$ and $y$ at the $z=0$ plane
 for a toroidal condensate with $ a=20\,a_B$.} 
\label{pot20}
\end{figure}

\begin{figure}[h]
 \includegraphics[width=0.8\linewidth,angle=0]{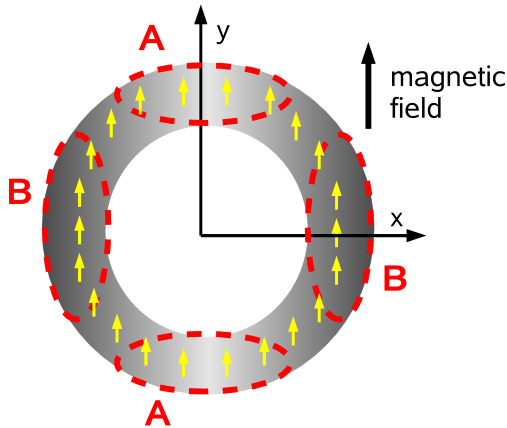}
\caption{(Color online) Diagram qualitatively showing the effective dipolar interaction in the torus.}
\label{FigTorus}
\end{figure}

To understand the difference between the two geometries 
we have drawn in Fig.~\ref{pot20} the mean-field dipolar potential for a condensate with $a=20\,a_B$ (compare with its corresponding harmonically trapped system in Fig.~\ref{pan20n}). We can see that it presents two deep minima on the $x$ axis and two saddle-points on the $y$ axis. The main consequence of this energy landscape is that the system will present attractive regions in the $x$ axis and repulsive regions in the $y$ axis, in contrast to the harmonic trap configuration. This situation 
can be easily understood
with the help of  Fig.~\ref{FigTorus}.
In this figure we have schematically delimited two types of regions A and B with dashed lines.
Within region A the dipoles are mainly displayed side-by-side, 
giving a net repulsive
interaction. In region B the atoms
are mainly located head-to-tail, which gives an effective attractive interaction. 
A straightforward effect of the location of regions A and B is that it is energetically favorable for the system to accommodate more dipoles in regions B. Therefore, a dipolar condensate confined in a toroidal trap shows large density maxima in the perpendicular direction to the magnetization axis, and small maxima (or saddle-points) in the magnetization direction. This is clearly seen in Fig.~\ref{torus14}, where in the central and right panels the density profiles along $x$, $y$ and $z$ are shown for condensates with $a=20\,a_B$ and $a=14\,a_B$, respectively. Note that in these cases the density along the $z$ axis (dotted line) is zero, since the Gaussian potential creates a hole along this axis. 
With these profiles in mind it is important to point out that the differences in the rms-radii (Table~\ref{tab:1}) 
 are mainly provided by the differences in the height of 
the density peaks and not
 by a major change in the size of the condensate itself,
which remains around the same value ($ \simeq 20\, \mu$m) in both directions.

\begin{figure}
\epsfig{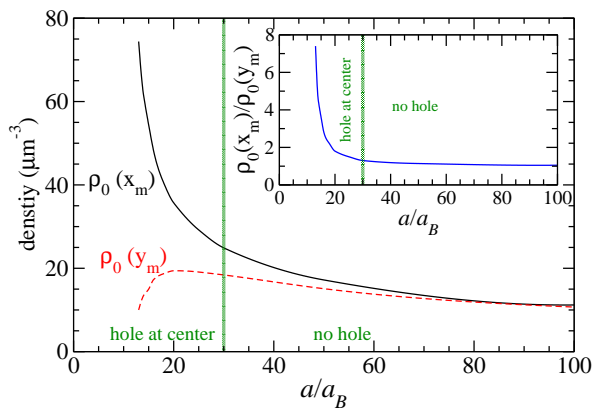}
\caption{(Color online) Density values at the critical points $ (x_m,0,0)$ and $(0,y_m,0)$,
which correspond to a maximum (solid line) and
a saddle point (dashed line), respectively. In the inset the quotient of these heights is drawn.} 
\label{denq}
\end{figure}

To quantitatively analyze the azimuthal density dependence, we show in Fig.~\ref{denq} the evolution of the maximum density values 
 along the $x$-axis,
 $ \rho_0(x_m) $,  and along the $y$-axis, $ \rho_0(y_m) $, as a function of the scattering length.
 A strong deviation of both quantities
occurs when the scattering length is reduced below the value in which the condensate geometry becomes multiply connected. The vertical line in Fig.~\ref{denq} marks the scattering length below which a hole in the condensate density appears.

\begin{figure}\centering\vspace{-1.5cm}
\epsfig{file=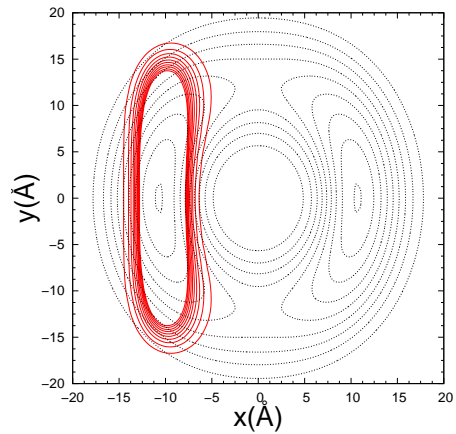,width=0.7\linewidth,clip=true}\vspace{-1cm}
\caption{(Color online) Density contours for the
 ground state configuration in the $(x,y)$ plane for a condensate with $ a=20\,a_B$ (dotted curves)
and $a=12\,a_B$ 
(solid curves). }
\label{dens1220}
\end{figure}

For an even smaller value of the scattering length, below $ a=13\,a_B$, a symmetry breaking
(SB) occurs.
The atoms are accumulated only around one of the peaks. 
We have checked this result by starting
 the minimization
procedure with different initial wave functions, including a symmetric wave function and a randomly generated one.
In all cases we have obtained the same final asymmetric state. 
In Fig.~\ref{dens1220} we compare the equidensity lines in the $z=0$ plane of two different configurations with $a=20\,a_B$ and $a=12\,a_B$. For $a=20\,a_B$  the system still presents reflection symmetry but for  $a=12\,a_B$ the symmetry is broken. It is interesting to observe that not only do the 
atoms accumulate at one
side of the condensate but also their density distribution  appears  more resistive to curve. 
This effect is a consequence of the fact that the dipolar 
interaction forces the
particles to locate themselves head-to-tail.

The symmetry breaking phenomenon presented here is a result of the combination of the toroidal geometry of the external potential and the anisotropic character of the mean-field dipolar interaction, which yields a self-induced energy barrier in the azimuthal direction (see Fig.~\ref{pot20}). 
When the scattering length is small, the attractive part of the dipolar interaction becomes large enough to be energetically favorable for the system to distribute all atoms in only one of the minima of the potential, which yields a symmetry broken configuration. With the present choice of parameters, a further reduction of the scattering length leads to the system collapse.

Recently, SB phenomena in dipolar condensates confined in double-well traps have been reported in Refs.~\cite{xi09} and \cite{Asad2009}. In Ref.~\cite{xi09}, the magnetization direction is used to induce SB, while in Ref.~\cite{Asad2009} it is the number of dipoles in the double well that drives it. The mechanism triggering the effect is the presence of a repulsive barrier in systems that are dominated by attractive interactions.
The main difference between the present configuration and the ones exposed in Refs.~\cite{xi09, Asad2009} is that we deal with a self-induced SB, since it is the dipolar interaction itself and its anisotropic character what brings about the effect. In contrast,
in the double-well case, SB should be observed in purely $s$-wave condensates with attractive interactions, provided the number of atoms is smaller than the critical one.

\section{Vortex states}\label{Vortex}

Toroidal traps have been experimentally shown to be capable of sustaining
 persistent flows \cite{ryu07}. This fact is related to the existence of 
metastable vortices \cite{cap09}.
In this section we study and characterize the anisotropic velocity flow that a phase-imprinted metastable vortex introduces in the toroidal condensate in presence of 
 dipolar interactions.

\subsection{Numerical generation  of vorticity and vortex density profiles}\label{Vortex1}

To generate vortex states along the $z$ axis we
 have used an ansatz already described in Ref. \cite{Jezek08},
namely we first obtain the ground state order parameter $\psi_0$ from Eq.~(\ref{gp}) and then 
imprint a velocity field in 
the following form:
\begin{equation}
\psi({\bf r})=\psi_0({\bf r}) \,\left( \frac{x+iy}{\sqrt{x^2+y^2}}\right)^n \,,
\label{wf-ansatz}
\end{equation}
with $n$ the quantum number related to the velocity circulation. We use this ansatz as the initial 
wave function of the imaginary-time evolution of the  GP equation (\ref{gp}). 
After convergence, we obtain the $n$-vortex state of the system. 
 Note that with our choice of magnetization axis 
($y$) the axial symmetry is removed and thus
the  angular momentum along the  $z$ axis is no longer $ L_z = n \hbar N $, since the 
angular momentum operator
does not commute with the Hamiltonian anymore.

A signal that the system can sustain 
metastable vortices \cite{cap09} is the presence of a valley in the 
ground state density landscape, which may produce a local vortex energy minimum.
In this case, after the minimization process, one can obtain a vortex state captured in the
toroidal trap due to a vortex energy barrier produced by the surrounding density 
maxima \cite{cap09}.
In our system, where the axial symmetry is not achieved, 
  the  energy barrier \cite{cap09} is related to the height of the saddle points 
 of the density lansdcape which are located  along the $y$-axis. 

\begin{figure}
\vspace*{0.5cm}
\epsfig{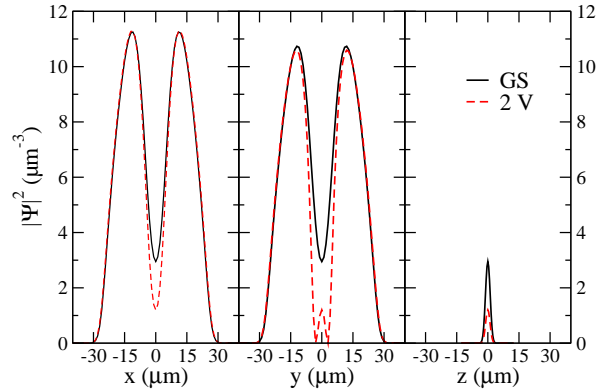}
\caption{(Color online) Density profiles for the
 ground state (solid line) and  two-vortex  state (dashed line) configurations along 
 $x$-axis,
$y$-axis, 
 and $z$-axis (left, middle, and central panels, respectively) for a condensate with $a=100\,a_B$.
}
 \label{as1002v}
\end{figure}

\begin{figure}
\epsfig{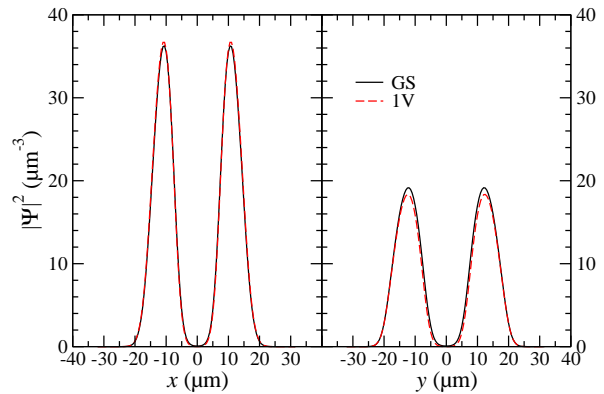}
\caption{(Color online) Density profiles for the
 ground state (solid line) and a single vortex state (dashed line) configurations 
along  $x$-axis  and $y$-axis (left and right panels, respectively) for a condensate with $a=20\,a_B$.
\label{as201v}
 }
\end{figure}

In Fig.~\ref{as1002v} we show the ground and two-vortex state densities for 
a dipolar condensate with $a=100\,a_B$. 
Since
the ground state  is a simply connected condensate, the density does not exhibit a hole,
 hence we find that multiply
quantized vortices are less favored than multiple singly-quantized vortices. 
Although the initial imprinting method 
produces
a doubly-quantized vortex along the $z$-axis at $ x=0$ and $y=0$, 
during the minimization process the
vortex splits into two singly-quantized vortices. Since the density maximum (saddle-point)
 is smaller along the $y$ direction the final state converges to two vortices placed at the 
$y$-axis (see central panel of Fig.~\ref{as1002v}). This stationary configuration is possible because the velocity field due to the other
vortex cancels with the contribution to the vortex velocity 
provided by the density inhomogeneity \cite{jeba08}.

We have found that the initial doubly-quantized vortex always splits in two singly-quantized vortices, even in the case of multiply connected geometry, i.e. with zero density at the center of the trap. However, for scattering lengths between $a=25\,a_B$ and $a=30\,a_B$ the density at the center of the trap is very small, since the vortices are very close together. Below $a=25\,a_B$ the system is not capable of sustaining two vortices, since the height at the saddle-points is not large enough.

To characterize the velocity field in the case of large asymmetric density configurations in the toroidal trap, we have studied singly quantized vortices. The velocity field will be analyzed in the next paragraph, but here we comment on the density distribution of such a configuration. In  Fig.~\ref{as201v} we show the density profiles for the ground state and a single-vortex state for a condensate with a scattering length $a=20\,a_B$. Note that along the $z$-axis both densities vanish. 
In this case, in contrast to the case of a large scattering length, the presence of the vortex does not produce a large change in the density distribution compared to the vortex-free configuration.

\subsection{Vortex velocity field}\label{Vortex2}

The initially imprinted velocity field, Eq.~(\ref{wf-ansatz}), of a single vortex corresponds to
a homogeneous medium or an axially symmetric condensate. 
Our system is not axially symmetric and
thus the modulus of  velocity field, for a given radius, varies around the torus.
This is simply understood by reminding that in stationary conditions the current intensity is constant along
the torus, 
\begin{equation}
I=\int {\bf j} \cdot d  \boldsymbol{\mathcal{S}} = 
\int |\psi|^2 \,{\bf v}({\bf r})\cdot d \boldsymbol{\mathcal{S}}\ ,
\label{current}
\end{equation}
where $ d \boldsymbol{\mathcal{S}}$ is the surface element. The velocity field $\mathbf{v}({\bf r})$ is related to the phase $S(\boldsymbol{r})$ 
of the wave function through
\begin{equation}
 {\bf v} ({\bf r})=\frac{\hbar}{m}\boldsymbol{\nabla} S({\bf r})\ .
\end{equation}
The integral in Eq.~(\ref{current}) is calculated in a torus section. It is easy to see from this equation that for the angles where
the density is lower (regions A) the velocity is larger, while in the angles where the density shows a maximum (regions B) the velocity is minimum.

\begin{figure}
\epsfig{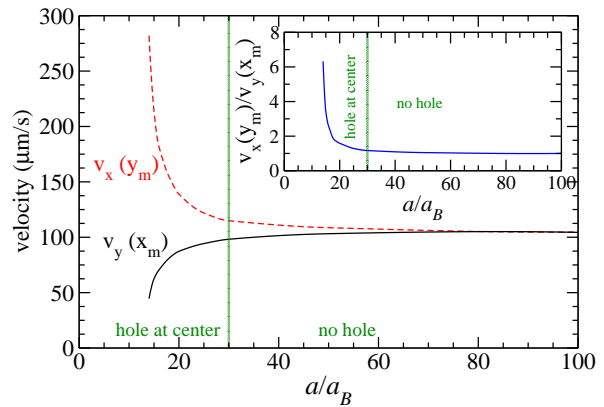}
\caption{(Color online) Velocity field components $v_y$ (solid line) and $v_x$ (dashed line) at the density maxima $x_m$ and $y_m$, respectively, as a function of the scattering length. In the inset, the ratio between the two is shown.}
 \label{velo}
\end{figure}

\begin{figure}\centering
\begin{minipage}{0.4\textwidth}\hspace{-2cm}\includegraphics[width=0.8\textwidth]{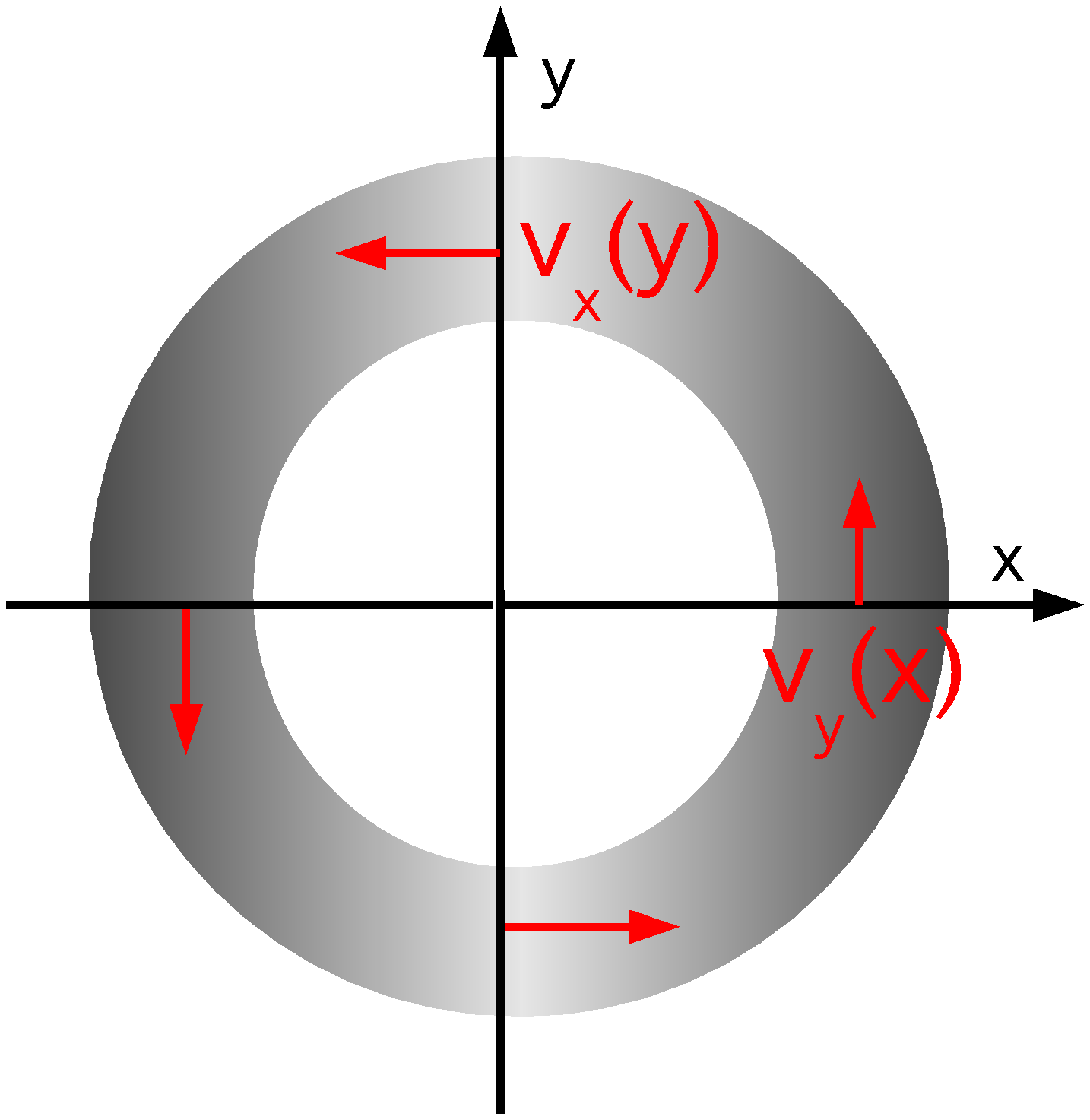}\end{minipage}%
\begin{minipage}{0.4\textwidth}\hspace{-9cm}\includegraphics[width=0.6\textwidth]{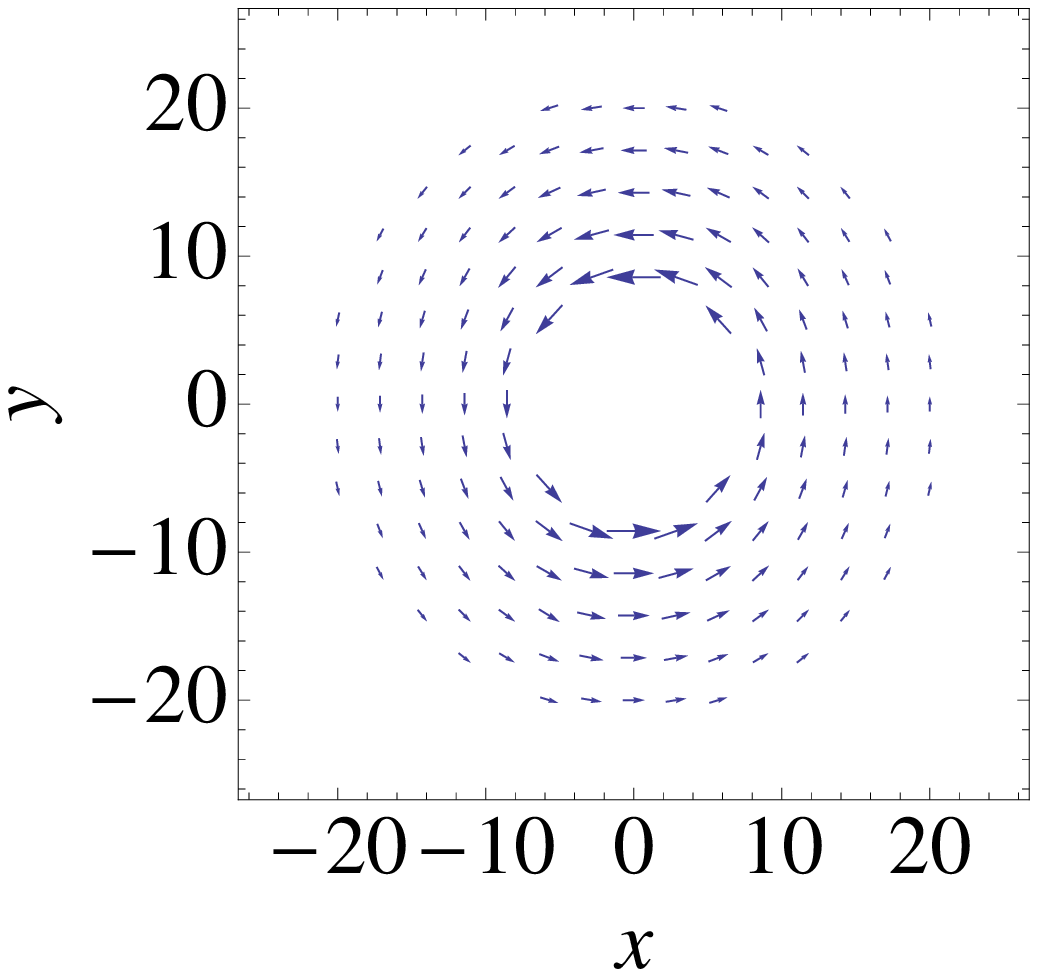}\end{minipage}
\caption{(Color online) Left panel: schematic diagram representing the components of the current 
from which the velocities depicted in Fig.~\ref{velo} are taken. Right panel: velocity field on the plane $z=0$ for $a=20\,a_B$.
}\label{esquemavelo}
\end{figure}

In Fig.~\ref{velo} we display the $x$ and $y$ components
of the velocity field at the positions $y_m$ and $x_m$ of the density maxima, respectively $v_x(y_m)$ and $v_y(x_m)$ (see also Fig.~\ref{esquemavelo}). For $ a=100\,a_B$, since the dipolar effects are small the densities along each axis are similar and, therefore, both
components of the velocity are equal. 
When we reduce the value of the scattering length, dipolar effects become sizeable and the density difference increases (see Fig.~\ref{denq}), hence decreasing the difference in both components of the velocity.
This can be used to generate a desired local velocity field intensity by only
tuning the scattering length using a Feshbach resonance. 

\begin{figure}
\vspace*{0.5cm}
\epsfig{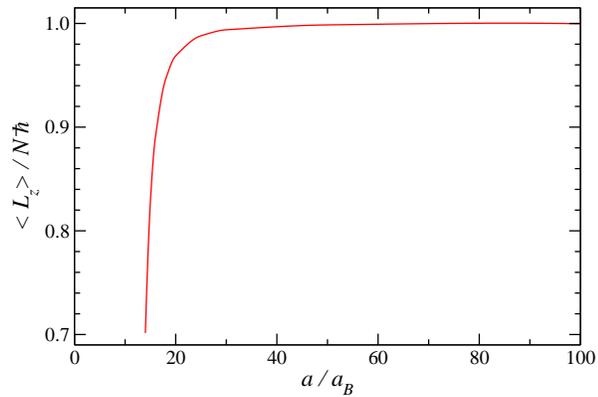}
\caption{Angular momentum as a function of the scattering length.
 }
 \label{lz}
\end{figure}

Although the angular momentum $L_z$ is not conserved, its expectation value $\langle L_z\rangle$ can provide a qualitative estimation of the difference between the velocity field in the different regions of the torus. 
As expected, for large scattering lengths the angular momentum per particle
 is almost one, see Fig.~\ref{lz}. As $a$ decreases the angular momentum decreases, 
being the variation much
stronger when the condensate exhibits a hole, below $a = 30\,a_B$. 
The reduction in the angular momentum is an evidence of the presence of self-induced energy barriers, which diminish the net particle flow. Since in our case the azimuthal energy barrier is larger when the scattering length is reduced, a measure of the angular momentum in the $z$ direction could provide information of the strength of the dipolar interaction with respect to the contact interaction.

\section{Summary and concluding remarks }\label{Conclusions}

In this paper we have studied the effects induced by the anisotropic nature of the dipolar interaction in a Bose-Einstein condensate of chromium atoms, especially in the case of toroidal trapping. We have considered the magnetization axis to be perpendicular to the harmonic trap symmetry axis and we have analyzed the system for different scattering lengths.

We have studied the system confined in a harmonic potential. In this case it is possible to tune its geometry just by varying the scattering length: for a large scattering length, the system shows mainly an oblate (pancake-shaped) geometry, while for small scattering lengths it can reach the configuration of a prolate (cigar-shaped) condensate.

The anisotropic character of the dipolar interaction can be further enhanced if a toroidal trap is used. The density then becomes inhomogeneously distributed along the torus, showing large and small density peaks as the azimuthal angle is varied for a fixed radius. The larger density peaks are observed in the $x$-axis, where the microscopic dipole-dipole interaction is repulsive but the net mean-field dipolar interaction is mainly attractive.

The difference between these density peaks in the $x$ and $y$ axes increases
as the scattering length is reduced. 
For a low enough  value of the scattering length, which in our configuration lies below $a=13\,a_B$, a  dipolar-induced symmetry 
breaking phenomenon occurs: the atoms are concentrated along only one of the major density peaks,
 reminding thus of a cigar-shaped condensate parallel to the magnetization direction.

Since toroidal traps can sustain persistent flows we
have  numerically obtained metastable vortex states using a
minimization procedure
with an initial phase-imprinted wave function. For the
geometry of the systems studied in the present work,
the velocity fields  show a strong azimuthal dependence.
Following a given constant radius along the torus, the absolute
value of the velocity is   smaller (larger)
where the density shows a maximum (minimum), in the intersection with
the $x$ axis ($y$ axis).
 We have also computed the expectation value of the angular momentum
along the trap symmetry axis. As we have chosen the dipolar interaction
direction perpendicular to this axis, the $z$ component of the angular momentum
 is not quantized.
Its reduction with respect to the non-dipolar  system
is an evidence of
the presence of self-induced energy barriers, which diminish
the net particle flow. Since the current intensity 
is conserved
along the torus, the net consequence of this
reduction is a larger difference
in magnitude of the velocity at the density critical points.

In conclusion we have found that the combination of dipolar interactions and toroidal trapping potentials leads to sizeable changes in the density and velocity field distributions 
that could be in principle experimentally observed.

\acknowledgments This work has been performed under Grants No. FIS2008-00421 from MEC (Spain), No. 2009SGR1289 from Generalitat de Catalunya (Spain),
and No. PICT 31980/05 from ANPCYT (Argentina). 
M. A. is supported by the Comission for Universities and Research of the Department of Innovation,
 Universities and Enterprises of the Catalan Government and the European Social Fund.


\begin{thebibliography}{99}

\bibitem{Werner2005} J. Werner, A. Griesmaier, S. Hensler, J. Stuhler, and T. Pfau, Phys. Rev. Lett. {\bf 94}, 183201 (2005).

\bibitem{Lahaye2009} T. Lahaye, C. Menotti, L. Santos, M. Lewenstein, and T. Pfau, Rep. Prog. Phys. \textbf{72}, 126401 (2009).

\bibitem{Stuhler2005} J. Stuhler, A. Griesmaier, T. Koch, M. Fattori, 
T. Pfau, S. Giovanazzi, P. Pedri, and L. Santos, 
Phys. Rev. Lett. \textbf{95}, 150406 (2005).

\bibitem{Giovanazzi2006} S. Giovanazzi, P. Pedri, L. Santos, A. Griesmaier, M. Fattori, T. Koch, J. Stuhler, and T. Pfau, Phys. Rev. A \textbf{74}, 013621 (2006).

\bibitem{Santos2000}
L. Santos, G. V. Shlyapnikov, P. Zoller, and M. Lewenstein,
Phys. Rev. Lett. {\bf 85}, 1791 (2000).

\bibitem{Koch2008}
T. Koch, T. Lahaye, J. Metz, B. Fr\"ohlich, A. Griesmaier, and
T. Pfau,
Nature Physics {\bf 4}, 220 (2008).

\bibitem{Eberlein2005}
C. Eberlein, S. Giovanazzi, and D. H. J. O'Dell,
Phys. Rev. A {\bf 71}, 033618 (2005).

\bibitem{Goral2002}
K. G\'oral and L. Santos, Phys. Rev. A {\bf 66}, 023613 (2002).

\bibitem{Yi2002} S. Yi and L. You, Phys. Rev. A \textbf{66}, 013607 (2002).

\bibitem{Dell2004} D. H. J. O'Dell, S. Giovanazzi, and C. Eberlein, Phys. Rev. Lett. \textbf{92}, 250401 (2004).

\bibitem{Ronen2006}
S. Ronen, D. C. E. Bortolotti and J. L. Bohn, Phys. Rev. A {\bf 74}, 013623 (2006).

\bibitem{Santos2003} 
L. Santos, G. Shlyapnikov, and M. Lewenstein, Phys. Rev. Lett. \textbf{90}, 250403 (2003).

\bibitem{Ronen2007}
S. Ronen, D. C. E. Bortolotti and J. L. Bohn, Phys. Rev. Lett. {\bf 98}, 030406 (2007); 
R. W. Wilson, S. Ronen, and J. L. Bohn,
 Phys. Rev. Lett. {\bf 100}, 245302 (2008).

\bibitem{Wilson2009a} 
R. M. Wilson, S. Ronen, and J. L. Bohn, Phys. Rev. A \textbf{80}, 023614 (2009).

\bibitem{Bijnen2007} 
C. Ticknor,  N. G. Parker, A. Melatos, S. L. Cornish, D. H. J. O'Dell, and A. M. Martin, Phys. Rev. A \textbf{78}, 061607(R) (2008); 
N. G. Parker, C. Ticknor, A. M. Martin, and D. H. J. O'Dell, Phys. Rev. A \textbf{79}, 013617 (2009).

\bibitem{Lahaye2008}
T. Lahaye, J. Metz, B. Fr\"{o}hlich, T. Koch, M. Meister, A. Griesmaier, T. Pfau, H. Saito, Y. Kawaguchi, and M. Ueda, Phys. Rev. Lett. \textbf{101}, 080401 (2008); J. Metz, T. Lahaye, B. Fr\"{o}hlich,  A. Griesmaier, T. Pfau, H. Saito, Y. Kawaguchi, and M. Ueda, New J. Phys. \textbf{11}, 055032 (2009).



\bibitem{gri05}
A. Griesmaier, J. Werner, S. Hensler, J. Stuhler, and T. Pfau,
 Phys. Rev. Lett. {\bf 94}, 160401 (2005).

\bibitem{Beaufils2008} 
Q. Beaufils, R. Chicireanu, T. Zanon, B. Laburthe-Tolra, E. Maréchal, L. Vernac, J.-C. Keller, and O. Gorceix, Phys. Rev. A \textbf{77}, 061601(R) (2008).

\bibitem{xi09} 
B. Xiong, J. Gong. H. Pu, W. Bao, and B. Li, Phys. Rev. A \textbf{79}, 013626 (2009).

\bibitem{Asad2009} 
M. Asad-uz-Zaman and D. Blume, Phys. Rev. A \textbf{80}, 053622 (2009).

\bibitem{ryu07} C. Ryu, M. F.  Andersen, P. Clad\'e, Vasant Natarajan,
 K. Helmerson,  and  W. D. Phillips, Phys. Rev. Lett. {\bf 99} 260401 (2007).

\bibitem{may08} T. Mayteevarunyoo, B. A. Malomed, and G. Dong,
 Phys. Rev. A \textbf{78}, 053601 (2008).


\bibitem{ol07} 
S. E. Olson, M. L Terraciano, M. Bashkansky, F. K. Fatemi, Phys. Rev. A \textbf{76}, 061404 (2007).

\bibitem{don91}
R. J. Donnelly, {\it Quantized Vortices in Helium II}
(Cambridge University Press, Cambridge, 1991).

\bibitem{Fetter2009} 
A. L. Fetter, Rev. Mod. Phys. \textbf{81}, 647 (2009).

\bibitem{wei08} C. N.  Weiler, T. W. Neely, D. R.  Scherer,  A. S. Bradley, M. J. Davis, and B. P. Anderson, Nature \textbf{455} 948 (2008).

\bibitem{cap09}
P. Capuzzi and D. M. Jezek, J. Phys. B {\bf 42}, 145301 (2009).

\bibitem{cat09}
H. M. Cataldo and D. M. Jezek, Eur. Phys. J. D {\bf 54}, 585 (2009).

\bibitem{mas09} 
P. Mason and N. G. Berloff,  Phys. Rev. A \textbf{79}, 043620 (2009).

\bibitem{pia09} 
F. Piazza, L. A. Collins, and A. Smerzi,  Phys. Rev. A \textbf{80}, 021601 (2009).


\bibitem{Yi2006}
S. Yi and H. Pu, Phys. Rev. A {\bf 73}, 061602(R) (2006).

\bibitem{dell07}
D. H. J. O'Dell and C. Eberlein, Phys. Rev. A  {\bf 75}, 013604 (2007).

\bibitem{Klawunn}
M. Klawunn, R. Nath, P. Pedri, and L. Santos, Phys. Rev. Lett. \textbf{100}, 240403 (2008); 
M. Klawunn and L. Santos, New J. Phys. \textbf{11} (5), 055012 (2009). 

\bibitem{Wilson2009} R. M. Wilson, S. Ronen, and J. L. Bohn, Phys. Rev. A \textbf{79}, 013621 (2009).

\bibitem{abad09} M. Abad, M. Guilleumas, R. Mayol, M. Pi, and D. M. Jezek, Phys. Rev. A \textbf{79}, 063622 (2009).

\bibitem{Jezek08}
D. M. Jezek, P. Capuzzi, and H. M. Cataldo, J. Phys. B {\bf 41}, 045304 (2008).

\bibitem{jeba08} 
D. M. Jezek,  P. Capuzzi, M. Guilleumas, and R. Mayol,   Phys. Rev. A \textbf{78}, 053616  (2008).





















\end{thebibliography}
\end{document}